\documentclass{article}

\usepackage{amssymb}
\usepackage{amsmath}
\usepackage{amsfonts}
\usepackage{mathrsfs}
\usepackage{hyperref}
\usepackage{amsthm}

\newtheorem{theorem}{Theorem}

\usepackage{graphicx}

\begin{document}

%
\noindent \textbf{\Large Axisymmetric Stationary  Spacetimes of  Constant  Scalar Curvature in Four Dimensions}
%

\vskip 1cm

\noindent {\large Rosikhuna F. Assafari}${}^{\sharp, a)}$, {\large Emir S. Fadhilla}${}^{\sharp, b)}$,  {\large Bobby E. Gunara}${}^{\S, \sharp, c)}$ \footnote{Corresponding author.}, {\large Hasanuddin}${}^{\ddagger, d}$, and {\large Abednego Wiliardy }${}^{\sharp, e)}$

\vskip 0.3cm

\noindent $^{\S}$\textit{ Indonesia Center for Theoretical and
Mathematical Physics (ICTMP)}
\\ { and} \\
$^{\sharp}$\textit{ Theoretical Physics Laboratory}\\
\textit{ Theoretical High Energy Physics and Instrumentation Research Group,}\\
\textit{ Faculty of Mathematics and Natural Sciences,}\\
\textit{ Institut Teknologi Bandung}\\
\textit{Jl. Ganesha no. 10 Bandung, Indonesia, 40132}
\\ { and} \\
$^{\ddagger}$\textit{ Department of Physics,}\\
\textit{ Faculty of Mathematics and Natural Science,}\\ 
\textit{ Tanjungpura University, }\\
\textit{ Jl. Prof. Dr. H. Hadari Nawawi, Pontianak, Indonesia, 78124 }

\noindent ${}^{a)}$rosikhuna@yahoo.co.id, ${}^{b)}$: emirsyahreza@students.itb.ac.id, ${}^{c)}$bobby@itb.ac.id, ${}^{d)}$hasanuddin@physics.untan.ac.id,  ${}^{e)}$abednego.wiliardy@gmail.com

\vskip  1cm

\begin{abstract}
In this paper we construct a special class  of four dimensional  axisymmetric stationary spacetimes whose Ricci scalar is constant but not Einstein. We find that this solution has a ring singularity. At the end, we discuss some numerical results of these spacetimes.
\end{abstract}

\vskip  1cm

\section{INTRODUCTION }

A family of four dimensional stationary axissymmetric spacetimes was firstly  derived by R. Kerr in 1963 describing uncharged rotating black holes which is a natural extension of static spacetimes called Schwarzschild solutions, see for example \cite{Chandrasekhar:1985kt, Teukolsky:2014vca}. So far, this class of solutions has been used in astrophysics explaining such as  quasars and accreting stellar-mass black hole systems, see for example \cite{ Teukolsky:2014vca}. \\
Therefore, it is of interest to study this stationary axissymmetric spacetime which is the aim of this paper. Here, we  construct  a special class of four dimensional   axisymmetric stationary spacetimes of constant scalar curvature  in the  Boyer-Lindquist coordinates. First, we discuss the construction of Einstein spacetimes (or known to be Kerr-Einstein spacetimes) with non-zero cosmological constant  by solving a modified Ernst equation for non-zero cosmological constant. In other words, we re-derive the Carter's result in \cite{Carter:1973rla}. Then, we proceed to construct the spaces of constant scalar curvature which are not Einstein by modifying the previous result, namely, we add two additional functions to the metric functions  to have  a more general  form of the metric but it has the structure of polynomial, namely, it is a quartic polynomial with five independent constants. This new family of axisymmetric stationary spacetimes of constant scalar curvature  still admits ring singularity. \\
\indent The structure of the paper can be mentioned as follows. In section \ref{sec:Axis} we give a quick review on axisymmetric stationary spacetimes. Then, we begin our construction of Kerr-Einstein spacetimes in section \ref{sec:Eins}.  We discuss the construction of axisymmetric stationary spacetimes of constant scalar curvature in section \ref{sec:Zwei} and study its singularity structure. In section \ref{Num} we show the numerical result of singularity discussion in section \ref{sec:Zwei}. Finally, we conclude our main results in section \ref{conclusions}

\section{Axisymmetric Stationary Spacetimes: A Quick Review }
\label{sec:Axis}

Suppose we have a metric in the general form 
\begin{equation} \label{metricgen} 
ds^2=g_{\mu \nu }(x) dx^{\mu }dx^{\nu } ~ , 
\end{equation} 
defined on a four dimensional spacetimes ${\bf M^4}$, where $x^{\mu}$ parametrizes a local chart on ${\bf M^4}$ and $\mu, \nu = 0,....,3$.  Then, we simplify the case as follows. In a stationary axisymmetric spacetime, the time coordinate $t$ and the azimuthal angle $\varphi $ are  considered to be $x^0$ and $x^1$ respectively. A stationary axisymmetric metric is invariant under simultaneous transformations $t\to -t$ and $\varphi \to -\varphi $ which  yields
\begin{equation} \label{zerocommetric} 
g_{02}=g_{03}=g_{12}=g_{13}=0  ~ ,
\end{equation} 
and moreover, all non-zero metric components depend only on $x^2 \equiv r$ and $x^3  \equiv \theta$. The latter condition implies  $g_{23}=0$ and the metric (\ref{metricgen}) can be simplified into \cite{Chandrasekhar:1977kf, Chandrasekhar:1985kt}
\begin{equation} \label{axisymmetric} 
ds^2 = - e^{2\nu } dt^2 + e^{2\psi }{\left(d\varphi -\omega dt\right)}^2 + e^{2{\mu }_2} dr^2 + e^{2{\mu }_3} d\theta^2 \ , 
\end{equation} 
where $(\nu, \psi, \omega, \mu_2, \mu_3) \equiv ( \nu(r, \theta), \psi(r, \theta), \omega(r, \theta), \mu_2(r, \theta), \mu_3(r, \theta) )$. In the following we list the non-zero components of Christoffel symbol related to the metric (\ref{axisymmetric}):
%
%
\begin{eqnarray}
{{\Gamma }^0}_{02}={\nu,}_2 -\frac{1}{2}\omega {\omega ,}_2 ~ e^{2(\psi -\nu)} ~ , ~  {{\Gamma }^0}_{03}={\nu,}_3-\frac{1}{2}\omega {\omega ,}_3e^{2(\psi -\nu)} ~ ,  \nonumber
\end{eqnarray}
\begin{eqnarray}
  {{\Gamma }^0}_{12} = \frac{1}{2} {\omega ,}_2e^{2(\psi -\nu)} ~ , ~ {{\Gamma }^0}_{13}= \frac{1}{2}  {\omega ,}_3e^{2(\psi -\nu)} ~ ,  \nonumber
\end{eqnarray}
\begin{equation}
 {{\Gamma }^1}_{20}=-\omega \left({\psi ,}_2-{\nu,}_2\right)-\frac{1}{2}{\omega ,}_2(1+{\omega }^2e^{2\left(\psi - \nu\right)}) ~ ,  \nonumber
\end{equation}
\begin{equation}
 {{\Gamma }^1}_{12}={\psi ,}_2+\frac{1}{2}\omega {\omega ,}_2 ~ e^{2(\psi -v)} ~ , ~ {{\Gamma }^1}_{13}={\psi ,}_3+\frac{1}{2}\omega {\omega ,}_3 ~ e^{2(\psi - \nu)} ~ ,  \nonumber
 \end{equation}
 \begin{equation} \label{ChrisSym}
 {{\Gamma }^2}_{00}={\nu,}_2 ~ e^{2(v-{\mu }_2)}-\omega ({\omega ,}_2+\omega {\psi ,}_2)e^{2(\psi -{\mu }_2)} ~,
 \end{equation}
\begin{equation}
 {{\Gamma }^2}_{01}=\left(\frac{1}{2}{\omega ,}_2+\omega {\psi ,}_2\right)e^{2(\psi -{\mu }_2)} ~ , ~ {{\Gamma }^2}_{11}=-{\psi ,}_2e^{2(\psi -{\mu }_2)} ~ ,  \nonumber
 \end{equation}
\begin{equation}
 {{\Gamma }^2}_{22}={{\mu }_2,}_2 ~ , ~ {{\Gamma }^2}_{23}={{\mu }_2,}_3 ~ , ~  {{\Gamma }^2}_{33}=-{{\mu }_3,}_2 ~ e^{2({\mu }_2-{\mu }_3)} ~ ,   \nonumber
\end{equation}
\begin{equation}
{{\Gamma }^3}_{00}={\nu,}_3e^{2(\nu-{\mu }_3)}-\omega ({\omega ,}_3+\omega {\psi ,}_3)e^{2(\psi -{\mu }_3)} ~ ,   \nonumber
\end{equation}
\begin{equation}
{{\Gamma }^3}_{01}=\left(\frac{1}{2}{\omega ,}_3+\omega {\psi ,}_3\right)e^{2(\psi -{\mu }_3)}~ , ~ {{\Gamma }^3}_{11}=-{\psi ,}_3 ~ e^{2(\psi -{\mu }_3)} ~ ,  \nonumber
\end{equation}
\begin{equation}
{{\Gamma }^3}_{22}=-{{\mu }_2,}_3 ~ e^{2({\mu }_2-{\mu }_3)} ~ , ~ {{\Gamma }^3}_{23}={{\mu }_3,}_2 ~ , ~ {{\Gamma }^3}_{33}={{\mu }_3,}_3 ~ ,  \nonumber
\end{equation}
%
and the non-zero components of Ricci tensor:
\begin{eqnarray}  
R_{00} &=&  e^{2\left(\nu-{\mu }_2\right)} \left({{\nu,}_2,}_2+{v,}_2{\left(\psi + \nu-{\mu }_2+{\mu }_3\right),}_2-\frac{1}{2}{{\omega }^2,}_2e^{2\left(\psi - \nu \right)}\right)   \nonumber\\
&& +e^{2\left(\nu-{\mu }_3\right)} \left({{\nu ,}_3,}_3+{\nu ,}_3{\left(\psi + \nu +{\mu }_2-{\mu }_3\right),}_3-\frac{1}{2}{{\omega }^2,}_3e^{2\left(\psi - \nu \right)}\right)    \nonumber\\
&& - \omega e^{2\left(\psi -{\mu }_2\right)}  \left({{\omega ,}_2,}_2+{\omega ,}_2{\left(3\psi - \nu -{\mu }_2+{\mu }_3\right),}_2 \right)  \nonumber\\
&& - \omega e^{2\left(\psi -{\mu }_3\right)} \left({{\omega ,}_3,}_3+{\omega ,}_3{\left(3\psi - \nu +{\mu }_2-{\mu }_3\right),}_3\right)   \nonumber\\  
&& -{\omega }^2   e^{2\left(\psi -{\mu }_2\right)}\left({{\psi ,}_2,}_2+{\psi ,}_2{\left(\psi + \nu -{\mu }_2+{\mu }_3\right),}_2+\frac{1}{2}{{\omega }^2,}_2e^{2\left(\psi - \nu \right)}\right)  \nonumber\\
&& -{\omega }^2  e^{2\left(\psi -{\mu }_3\right)}\left({{\psi ,}_3,}_3+{\psi ,}_3{\left(\psi + \nu +{\mu }_2-{\mu }_3\right),}_3+\frac{1}{2}{{\omega }^2,}_3e^{2\left(\psi - \nu \right)}\right)   ~ , \nonumber\\
R_{01} &=&   \frac{1}{2} e^{2\left(\psi -{\mu }_2\right)}\left({{\omega ,}_2,}_2 + {\omega ,}_2{\left(3\psi - \nu -{\mu }_2+{\mu }_3\right),}_2\right) \nonumber\\
&& +\frac{1}{2} e^{2\left(\psi -{\mu }_3\right)}\left({{\omega ,}_3,}_3+{\omega ,}_3{\left(3\psi - \nu +{\mu }_2-{\mu }_3\right),}_3\right) \nonumber\\
&& + \omega   e^{2\left(\psi -{\mu }_2\right)}\left({{\psi ,}_2,}_2+{\psi ,}_2{\left(\psi + \nu -{\mu }_2+{\mu }_3\right),}_2+\frac{1}{2}{{\omega }^2,}_2e^{2\left(\psi - \nu \right)}\right)  \nonumber\\
&& +   \omega e^{2\left(\psi -{\mu }_3\right)}\left({{\psi ,}_3,}_3+{\psi ,}_3{\left(\psi + \nu +{\mu }_2-{\mu }_3\right),}_3+\frac{1}{2}{{\omega }^2,}_3e^{2\left(\psi - \nu \right)}\right)  ~ , \label{riccitensor} \\
\end{eqnarray}
\begin{eqnarray}
R_{11} &=& - e^{2\left(\psi -{\mu }_2\right)} \left({{\psi ,}_2,}_2+{\psi ,}_2{\left(\psi + \nu -{\mu }_2+{\mu }_3\right),}_2+\frac{1}{2}{{\omega }^2,}_2e^{2\left(\psi - \nu \right)}\right) \nonumber\\
&&  - e^{2\left(\psi -{\mu }_3\right)} \left({{\psi ,}_3,}_3   + {\psi ,}_3{\left(\psi + \nu +{\mu }_2-{\mu }_3\right),}_3+\frac{1}{2}{{\omega }^2,}_3 e^{2\left(\psi - \nu \right)}\right) ~ , \nonumber\\
R_{22} &=& -({{\psi ,}_2,}_2+{\psi ,}_2{\left(\psi -{\mu }_2\right),}_2-({{ \nu ,}_2,}_2+{v,}_2{\left( \nu -{\mu }_2\right),}_2 \nonumber\\
&& -e^{2\left({\mu }_2-{\mu }_3\right)}\left({\mu }_{2,3,3}+{\mu }_{2,3}{\left(\psi + \nu +{\mu }_2-{\mu }_3\right),}_3\right) \nonumber\\
&& - \left({\mu }_{3,2,2}+{\mu }_{3,2}{\left({\mu }_3-{\mu }_2\right),}_2\right)+\frac{1}{2}{{\omega }^2,}_2 e^{2\left(\psi -v\right)} ~ , \nonumber\\ 
R_{23} &=& -({{\psi ,}_2,}_3+{\psi ,}_2{\left(\psi -{\mu }_2\right),}_3-({{\nu ,}_2,}_3+{\nu ,}_2{\left(v-{\mu }_2\right),}_3+{\mu }_{3,2}{\left(\psi - \nu\right),}_3 \nonumber\\
&& +\frac{1}{2}{\omega ,}_2{\omega ,}_3e^{2\left(\psi - \nu \right)} ~ , \nonumber\\ 
R_{33} &=& -({{\psi ,}_3,}_3+{\psi ,}_3{\left(\psi -{\mu }_3\right),}_3-({{\nu ,}_3,}_3 + {\nu ,}_3{\left(v-{\mu }_3\right),}_3 \nonumber\\
&& - e^{2\left({\mu }_3-{\mu }_2\right)}\left({\mu }_{3,2,2}+{\mu }_{3,2}{\left(\psi + \nu -{\mu }_2+{\mu }_3\right),}_3\right)\nonumber\\
&& - \left({\mu }_{2,3,3}+{\mu }_{2,3}{\left({\mu }_2-{\mu }_3\right),}_3\right)+\frac{1}{2}{{\omega }^2,}_3 e^{2\left(\psi - \nu \right)} ~ . \nonumber
\end{eqnarray} 
Then,  Ricci scalar can be obtained as
\begin{eqnarray} 
- R &=& 2e^{-2{\mu }_2} \Big(  {\psi }_{,2,2}+{\psi }_{,2}{\left(\psi -{\mu }_2+{\mu }_3\right)}_{, 2} +{\psi }_{, 2} {\nu ,}_2+ \nu_{,2,2}+{\nu }_{, 2} {\left(\nu  -{\mu }_2+{\mu }_3\right)}_{, 2} \nonumber\\ 
&& +{\mu }_{3,2,2}+{\mu }_{3,2}{\left({\mu }_3-{\mu }_2\right),}_2-\frac{1}{4}{{\omega }^2}_{, 2} e^{2\left(\psi - \nu \right)} \Big) \nonumber\\ 
&& + 2e^{-2{\mu }_3}  \Big(  {\psi }_{,3,3}+{\psi }_{, 3}{\left(\psi +{\mu }_2-{\mu }_3\right),}_3+{\psi}_{, 3} {\nu }_{, 3} +  \nu_{,3,3}+ {\nu}_{, 3} {\left(\nu +{\mu }_2-{\mu }_3\right),}_3 \nonumber\\ 
&& +{\mu }_{2,3,3}+{\mu }_{2,3}{\left({\mu }_2-{\mu }_3\right)}_{, 3} -\frac{1}{4}{{\omega }^2}_{, 3} e^{2\left(\psi - \nu \right)}  \Big) ~ ,   \label{ricciscalar} 
\end{eqnarray} 
where we have defined
\begin{equation} 
{f }_{,\mu} \equiv \frac{\partial f}{\partial x^{\mu}} ~ , \quad {f }_{, \mu , \nu} \equiv \frac{\partial^2 f}{\partial x^{\mu} \partial x^{\nu} } ~ .
\end{equation}

\section{EINSTEIN SPACETIMES}
\label{sec:Eins}

In this section, we construct a class of axisymmetric spacetimes satisfying Einstein condition
\begin{equation} \label{Einsteincon} 
R_{\mu \nu } = \Lambda g_{\mu \nu } ~ ,
\end{equation} 
with $\Lambda$ is named cosmological constant, yielding the following coupled nonlinear equations:
\begin{equation} \label{Einsteincon1} 
{\left(e^{{\rm 3}\psi {\rm -} \nu{\rm -}{\mu }_{{\rm 2}}{\rm +}{\mu }_{{\rm 3}}}{\omega}_{, 2} \right)}_{, 2} {\rm +}{\left(e^{{\rm 3}\psi {\rm -} \nu {\rm +}{\mu }_{{\rm 2}}{\rm -}{\mu }_{{\rm 3}}}{\omega }_{, 3} \right)}_{ , 3} = 0 ~ , 
\end{equation}
\begin{equation} \label{Einsteincon2} 
{(\psi + \nu)}_{,2,3}-{\left(\psi + \nu \right)}_{, 2}{\mu }_{2,3}-{\left(\psi + \nu \right)}_{,3}{\mu }_{3,2}+{\psi }_{,2}{\psi }_{,3}+ \nu_{,2} \nu_{,3}=\frac{1}{2} e^{2\left(\psi - \nu \right)}{\omega }_{, 2} ~ {\omega }_{, 3}  ~ , 
\end{equation} 
%
%
\begin{equation} \label{Einsteincon3} 
{\left(e^{{\mu }_3-{\mu }_2} {\left(e^{\beta }\right)}_{, 2} \right)}_{, 2} +{\left(e^{{\mu }_2-{\mu }_3}{\left(e^{\beta }\right)}_{, 3} \right)}_{, 3} = -2\Lambda e^{\beta +{\mu }_2+{\mu }_3} ~ , 
\end{equation} 
%
\begin{equation} \label{Einsteincon4}  
{\left(e^{\beta -{\mu }_2+{\mu }_3}{(\psi - \nu)}_{, 2} \right)}_{, 2} + {\left(e^{\beta +{\mu }_2-{\mu }_3}{(\psi - \nu)}_{, 3}  \right)}_{, 3} = -e^{3\psi - \nu}\left(e^{{\mu }_3-{\mu }_2}{{\omega}_{, 2}}^2 + e^{{\mu }_2-{\mu }_3}{{\omega}_{, 3}}^2\right), 
\end{equation} 
%
\begin{eqnarray}
 {{\rm 4}e}^{{\mu }_{{\rm 3}}{\rm -}{\mu }_{{\rm 2}}}\left({\beta ,}_{{\rm 2}}{\mu }_{{\rm 3,2}}{\rm +}{\psi}_{, 2} {\nu}_{, 2}\right) {\rm -}{{\rm 4}e}^{{\mu }_{{\rm 2}}{\rm -}{\mu }_{{\rm 3}}}\left({\beta}_{, 3}{\mu }_{{\rm 2,3}}{\rm +}{\psi }_{, 3} {\nu}_{, 3}\right)  &=& {{\rm 2}e}^{{\rm -}\beta }\left[{\left(e^{{\mu }_{{\rm 3}}{\rm -}{\mu }_{{\rm 2}}}{\left(e^{\beta }\right),}_{{\rm 2}}\right)}_{, 2 }{\rm +}{\left(e^{{\mu }_{{\rm 2}}{\rm -}{\mu }_{{\rm 3}}}{\left(e^{\beta }\right)}_{, 3}\right)}_{, 3} \right] \nonumber\\
&& {\rm -}e^{{\rm 2}\left(\psi {\rm -} \nu \right)}\left(e^{{\mu }_{{\rm 3}}{\rm -}{\mu }_{{\rm 2}}}{{\omega}_{, 2}}^{{\rm 2}}{\rm -}e^{{\mu }_{{\rm 2}}{\rm -}{\mu }_{{\rm 3}}}{{\omega}_{, 3}}^{{\rm 2}}\right) ~  ,  \label{Einsteincon5}  
\end{eqnarray}  
where we have defined
\begin{equation}
\beta \equiv  \psi + \nu ~.
\end{equation} 
This class of solutions is called Kerr-(anti) de Sitter solutions.

\subsection{The Functions  $\mu_2$ and $\mu_3$}
First of all, we simply take $e^{\mu_2}$ as
\begin{equation} \label{emu2} 
e^{{\mu }_2}=\frac{(r^2+a^2{\cos }^2  \theta)^{\frac{1}{2}}}{\Delta^{(0)\frac{1}{2} }_r} ~ , 
\end{equation}
where $\Delta^{(0)}_r \equiv \Delta^{(0)}_r (r)$ and $a$ is a constant related to the angular momentum of a black hole \cite{Chandrasekhar:1985kt}. Next, we assume that the function $e^{2\left({\mu }_3-{\mu }_2\right)}$  and $e^{2\beta }$ are separable as
\begin{eqnarray} 
e^{2\left({\mu }_3-{\mu }_2\right)} &= &{\Delta }^{(0)}_r  \frac{{{\sin }^2 \theta \ }}{{\Delta }^{(0)}_{\theta } } ~ , \nonumber\\
e^{2\beta } &=& {\Delta }^{(0)}_r {\Delta }^{(0)}_{\theta }   ~ , \label{beta} 
\end{eqnarray}
with ${\Delta }^{(0)}_{\theta }  \equiv {\Delta }^{(0)}_{\theta }(\theta)$. Thus, (\ref{Einsteincon3}) can be cast into the form 
\begin{equation} \label{Einsteincon31} 
\left[ \Delta^{(0)\frac{1}{2} }_r {\left(\Delta^{(0)\frac{1}{2} }_r\right)}_{,2}\right]_{, 2 } +\frac{1}{{\sin  \theta \ }}  \left[\frac{{\Delta }^{(0)\frac{1}{2}}_{\theta }}{{\sin  \theta \ }}{\left({\Delta }^{(0)\frac{1}{2}}_{\theta }\right)}_{, 3 }\right]_{, 3 } = - 2\Lambda \left(r^2+a^2{{\cos }^{{\rm 2}} \theta \ }\right). 
\end{equation} 
Employing the variable separation method, we then obtain
\begin{eqnarray}
{\Delta }^{(0)}_r  &=&  - \frac{\Lambda}{3}  r^4 + c_1 r^2  + c_2 r + c_3 ~ , \nonumber\\
{\Delta }^{(0)}_{\theta } &=&  - \frac{\Lambda}{3} a^2 \cos^4 \theta - c_1 \cos^2 \theta  - c_4 \cos \theta + c_5 ~, \label{Deltasoleinsgeneral}
\end{eqnarray}
where $c_i \ i=1,...,5,$ are real constant. To make a contact with \cite{Carter:1973rla}, one has to set $c_i$ to be
\begin{eqnarray}
c_1 &=& -\frac{\Lambda}{3} a^2 ~ , ~ c_2 = -2M ~ , ~ c_3 = a^2 \  , \nonumber\\
c_4 &=& 0  ~ , ~ c_5 = 1 \  ,
\end{eqnarray}
such that we have 
\begin{eqnarray}
{\Delta }^{(0)}_r  &=& - \frac{\Lambda}{3}  r^2 \left(r^{{\rm 2}}{\rm +}a^{{\rm 2}}\right){\rm +}r^{{\rm 2}}{\rm -}{\rm 2}Mr{\rm +}a^{{\rm 2}} ~ , \nonumber\\
{\Delta }^{(0)}_{\theta }  &=& \left(1 + \frac{\Lambda}{3} a^2{{\cos }^2 \theta \ }\right) {\sin }^2 \theta   ~.  \label{Deltasoleins}
\end{eqnarray}
\subsection{The Functions $(\omega, \nu, \psi)$ and Ernst Equation}
To obtain the explicit form of $(\omega, \nu, \psi)$, we have to transform (\ref{Einsteincon1}) and (\ref{Einsteincon4}) into a so called  Ernst equation with non-zero $\Lambda$ using  (\ref{Deltasoleins}). This can be structured as follows.\\
\indent First, we introduce a pair of functions $(\Phi, \Psi)$ via
\begin{eqnarray}
\Phi_{, 2} &=&  e^{2(\psi - \nu)} {\Delta }^{(0)}_{\theta } \omega_{, p} ~ , \nonumber\\
\Phi_{, p} &=&  - e^{2(\psi - \nu)} {\Delta }^{(0)}_r \omega_{, 2} ~ , \\
\Psi &\equiv&  e^{\psi - \nu} \Delta^{(0)\frac{1}{2} }_r \Delta^{(0)\frac{1}{2} }_{\theta }  ~ ,\nonumber
\end{eqnarray}
where $p \equiv \cos \theta$. Then, (\ref{Einsteincon1}) and (\ref{Einsteincon4})  can be cast into
\begin{eqnarray}
\Psi \left[ ({\Delta }^{(0)}_r  \Phi_{, 2} )_{, 2} + ({\Delta }^{(0)}_{\theta}\Phi_{, p})_{, p}  \right]  &=&  2 {\Delta }^{(0)}_r  \Psi_{, 2}  \Phi_{, 2} + 2  {\Delta }^{(0)}_{\theta} \Psi_{, p}  \Phi_{, p}  ~ , \label{enrsteq}\\
\Psi \left[ ({\Delta }^{(0)}_r  \Psi_{, 2} )_{, 2} + ({\Delta }^{(0)}_{\theta}\Psi_{, p})_{, p}  \right]  &=& \Delta^{(0)}_r  \left[ (\Psi_{, 2} )^2 - (\Phi_{, 2} )^2  \right]  - \Delta^{(0)}_{\theta }  \left[ (\Psi_{, p} )^2 - (\Phi_{, p} )^2  \right]  ~ , \nonumber
\end{eqnarray}
respectively. Defining a complex function $Z \equiv \Psi  + {\mathrm i}\Phi $, (\ref{enrsteq}) can be rewritten in Ernst form
\begin{equation} 
{\mathrm Re}Z   \left[ ({\Delta }^{(0)}_r  Z_{, 2} )_{, 2} + ({\Delta }^{(0)}_{\theta}Z_{, p})_{, p}  \right]  = \Delta^{(0)}_r (Z_{, 2} )^2 + \Delta^{(0)}_{\theta }   (Z_{, p} )^2 ~ . \label{enrsteq1} 
\end{equation} 
Note that one could obtain another solution of (\ref{enrsteq1}), say $\tilde{Z} \equiv \tilde{\Psi}  + {\mathrm i} \tilde{\Phi}$ by a conjugate transformation
\begin{eqnarray}
\tilde{\Psi} &=& \frac{\Delta^{(0)\frac{1}{2} }_r \Delta^{(0)\frac{1}{2} }_{\theta} }{\tilde{\chi}} ~ , \nonumber\\
\tilde{\Phi}_{, 2} &=& \frac{\Delta^{(0) }_{\theta}}{\tilde{\chi}^2} \tilde{\omega} _{, 3}  ~ , \label{transcon}\\
\tilde{\Phi}_{, 3} &=& - \frac{\Delta^{(0) }_r}{\tilde{\chi}^2} \tilde{\omega} _{, 2}  ~,\nonumber
\end{eqnarray}
where
\begin{eqnarray}
\tilde{\chi} &\equiv&  \frac{e^{\nu - \psi}}{e^{2(\nu - \psi)}-\omega^2} ~ , \nonumber\\
\tilde{\omega}  &\equiv&   \frac{\omega}{e^{2(\nu - \psi)}-\omega^2} ~ .
\end{eqnarray}
In the latter basis, we find
\begin{eqnarray}
\tilde{\Psi} &=&  \frac{\Delta^{(0) }_r - a^2 \Delta^{(0) }_{\theta} }{ r^2+a^2 {\cos }^2 \theta } ~ , \nonumber\\
\tilde{\Phi} &=& \frac{2aM \cos \theta}{r^2+a^2{{\cos }^2 \theta}}  + \frac{2 \Lambda}{3} a r  \cos \theta ~ .
\end{eqnarray}
After some computation, we conclude that \cite{Carter:1973rla}
\begin{eqnarray}
e^{2\psi } &=& \frac{ \left(r^2+a^2\right)^2 \Delta^{(0) }_{\theta} - \Delta^{(0) }_r a^2 {\sin }^4 \theta } { r^2+a^2{{\cos }^2 \theta } }   ~ , \nonumber\\
e^{2\nu } &=&  \frac{( r^2+a^2 {\cos }^2 \theta ) \Delta^{(0) }_{\theta} \Delta^{(0) }_r }{  \left(r^2+a^2\right)^2 \Delta^{(0) }_{\theta} - \Delta^{(0) }_r a^2 {\sin }^4 \theta  } ~, \label{finalsoleins} \\
\omega &=& \frac{ a (r^2+a^2) \Delta^{(0) }_{\theta}  - a  \ {\sin }^2 \theta) \Delta^{(0) }_r } { \left(r^2+a^2\right)^2 \Delta^{(0) }_{\theta} - \Delta^{(0) }_r a^2 {\sin }^4 \theta }  ~ .\nonumber
\end{eqnarray}

\section{SPACETIMES OF CONSTANT RICCI SCALAR}
\label{sec:Zwei}

In this section we extend the previous results to the case of spaces of constant Ricci scalar, namely
\begin{equation} \label{riccicons} 
R=g^{\mu \nu }R_{\mu \nu } = k ~ ,
\end{equation} 
where $k$ is a constant.  To have an explicit solution, we simply replace $\Delta^{(0) }_r$ and $ \Delta^{(0) }_{\theta}$ in (\ref{beta}), (\ref{Deltasoleins}), and (\ref{finalsoleins}) by
\begin{eqnarray}
\Delta_r &=&  \Delta^{(0) }_r + f(r) ~ , \nonumber\\
\Delta_{\theta} &=& \Delta^{(0) }_{\theta}  + h(\theta) ~ .  \label{newdelta}
\end{eqnarray}
Then, inserting these modified functions mentioned above to (\ref{riccicons}), we simply have
\begin{equation} \label{newdeltaeq} 
- k ( r^2+a^2 {\cos }^2 \theta)  ={\left({\Delta }_r\right)}_{,2,2}+\frac{1}{{\sin  \theta \ }}{\left[\frac{{\left({\Delta }_{\theta }\right)}_{, 3 }}{{\sin  \theta \ }}\right]}_{, 3 } ~ ,
\end{equation} 
which gives
\begin{equation} \label{newdeltaeq1} 
- \left(k-4\Lambda \right) ( r^2+a^2 {\cos }^2 \theta) = 2{\left[f^{\frac{1}{2}}\ {\left(f^{\frac{1}{2}}\right)}_{, 2}\right]}_{, 2}+\frac{2}{{\sin  \theta \ }}{\left[\frac{h^{\frac{1}{2}}}{{\sin  \theta \ }}{\left(h^{\frac{1}{2}}\right)}_{, 3 }\right]}_{, 3 }. 
\end{equation} 
The solution of \eqref{newdeltaeq1} is given by
\begin{eqnarray} 
f(r) &=& - \frac{1}{12}\left(k-4\Lambda \right)r^4+\frac{1}{2}C_1r^2+C_2r+C_3 ~ , \nonumber\\
h(\theta)  &=&-  \frac{1}{12}\left(k-4\Lambda \right)a^2{{\cos }^4 \theta \ }-\frac{1}{2}C_1{{\cos }^2 \theta \ }-C_4{\cos  \theta \ }+C_5 ~ , \label{solfh}
\end{eqnarray}
where $C_i , \ i=1,...,5$, are real constant.  It is worth mentioning some remarks as follows. First, the functions $f(r)$ and $g(\theta)$ have the same structure as in (\ref{Deltasoleinsgeneral}), namely they are quartic polynomials with respect to $r$ and $\cos \theta$, respectively.  Second, the constant $\Lambda$ here is no longer the cosmological constant. Finally, Einstein spacetimes can be obtained by setting  $k = 4 \Lambda$  and $C_3 = a^2 C_5$ with other  $C_i$  \ i=1, 2, 4, are free constants.\\
\indent Now we can state our main result as follows.
\begin{theorem}\label{theor}
Suppose we have an axisymmetric spacetime ${\bf M^4}$ endowed with metric
\begin{equation}
ds^2= -e^{2\nu } dt^2 + e^{2\psi }{\left(d\varphi -\omega dt\right)}^2 + e^{2{\mu }_2} dr^2 + e^{2{\mu }_3} d\theta^2 \ , 
\label{axisymmetricsol} 
\end{equation} 
satisfying
\begin{eqnarray}
e^{2\left({\mu }_3-{\mu }_2\right)} &= &{\Delta }_r  \frac{{{\sin }^2 \theta \ }}{ {\Delta }_{\theta } } ~ , \nonumber\\
e^{2\beta } &=& {\Delta }_r {\Delta }_{\theta }   ~ , \nonumber\\
e^{2\psi } &=&  \frac{ \left(r^2+a^2\right)^2 \Delta_{\theta} - \Delta_r a^2 {\sin }^4 \theta } { r^2+a^2{{\cos }^2 \theta } }  ~ , \\
e^{2\nu } &=&  \frac{ ( r^2+a^2 {\cos }^2 \theta ) \Delta_{\theta} \Delta_r }{  \left(r^2+a^2\right)^2 \Delta_{\theta} - \Delta_r a^2 {\sin }^4 \theta  } ~,  \nonumber\\
\omega &=& \frac{ a (r^2+a^2) {\Delta }_{\theta }  - a \  {\sin }^2 \theta  \Delta_r} { \left(r^2+a^2\right)^2 \Delta_{\theta} - \Delta_r a^2 {\sin }^4 \theta }  ~ , \nonumber
\end{eqnarray}
where $\Delta_r$ and ${\Delta }_{\theta } $ are given by (\ref{newdelta}). Then, there exist a familiy of spacetimes of constant scalar curvature with
\begin{eqnarray}\label{sol}
\Delta_r &=&  - \frac{1}{3}\Lambda r^2 a^2+r^2-2 M r + a^2 -\frac{k}{12} r^4+\frac{1}{2} C_1r^2+C_2r+C_3 ~ , \\
\Delta_{\theta} &=& - {{\cos }^2 \theta \ } + \frac{\Lambda }{3} a^2{{\cos }^2 \theta \ } - \frac{k}{12} a^2{{\cos }^4 \theta \ }-\frac{1}{2}C_1{{\cos }^2 \theta \ }-C_4{\cos  \theta \ }+C_5  + 1 ~ , \nonumber  
\end{eqnarray}
where $C_i , \ i=1,...,5,$ are real constant.  The metric (\ref{axisymmetricsol}) becomes Einstein if   $k = 4\Lambda $ and $C_3 = a^2 C_5$.

\end{theorem}
\begin{proof}

Suppose $\quad f_{\mu\nu} = R_{\mu\nu} - \Lambda g_{\mu\nu}$, then for  the metric (\ref{axisymmetricsol}) we have 
\begin{eqnarray}
f_{00}&=&\frac{1}{4} g_{00} (k-4 \Lambda )+\frac{\left(C_{3}-a^2 C_{5}\right) \left(a^2 \Delta_{\theta }+\Delta_{r}\right)}{\rho ^6} ~ , \nonumber \\
f_{10}&=&\frac{1}{4} g_{10} (k-4 \Lambda )-\frac{\left(C_{3}-a^2 C_{5}\right) \left(a \left(\left(a^2+r^2\right) \Delta_{\theta }+\sin ^2(\theta ) \Delta_{r}\right)\right)}{\rho ^6} ~ , \nonumber \\
f_{11}&=&\frac{1}{4} g_{11} (k-4 \Lambda )+\frac{\left(C_{3}-a^2 C_{5}\right) \left(\left(a^2+r^2\right)^2 \Delta_{\theta }+a^2 \sin ^4(\theta ) \Delta_{r}\right)}{\rho ^6} ~ , \\
f_{22}&=&\frac{1}{4} g_{22} (k-4 \Lambda )-\frac{\left(C_{3}-a^2 C_{5}\right)}{\rho ^2 \Delta_{r}} ~ , \nonumber \\
f_{33}&=&\frac{1}{4} g_{33} (k-4 \Lambda )+\frac{\left(C_{3}-a^2 C_{5}\right) \sin ^2(\theta )}{\rho ^2 \Delta_{\theta }}  ~ , \nonumber
\end{eqnarray}
whereas the other components vanish. Then, the trace of $f_{\mu\nu}$ is given by
\begin{equation}
g^{\mu\nu} f_{\mu\nu} = k - 4 \Lambda ~ ,
\end{equation}
which implies that $R = k$.

\end{proof}

The norm of Riemann tensor for the case at hand in general has the form 
\begin{eqnarray}\label{Kretschmann}
R^{\mu\nu\alpha\beta}R_{\mu\nu\alpha\beta} 
&=& \frac{384 r^4 }{( r^2+a^2 {\cos }^2 \theta )^6} \Bigg(2 a^2 C_{4} \cos \theta  \Big(-a^2   C_5 + r (C_2 - 2 M)+C_3 \Big) \nonumber \\
&& + \Big(- a^2 C_5   + r (a C_4 +C_{2}-2 M)+C_3 \Big) \Big(- C_5 a^2 + r (-a C_4 + C_2 - 2 M)+C_3 \Big) \Bigg) \nonumber \\
&& +\frac{192 r^2 } {( r^2+a^2 {\cos }^2 \theta )^5}  \Bigg(a^2 C_{4} \cos \theta  \left(3 a^2 C_5 - 4 C_2 r - 3 C_3 + 8 M r \right) \nonumber \\
&& -5 r (C_{2}-2 M) \left(C_{3}-a^2 C_{5}\right)-2 \left(C_{3}-a^2 C_{5}\right)^2 \nonumber \\
&& - 3 r^2 (-a C_{4}+C_{2}-2 M) (a C_{4}+C_{2}-2 M)\Bigg) \nonumber \\
&& +\frac{8}{( r^2+a^2 {\cos }^2 \theta )^4} \Bigg(6 a^2 C_{4} \cos \theta  \Big(- a^2 C_5 + 3 r (C_{2}-2 M)+C_{3}\Big) \nonumber \\
&& + 30 r (C_{2}-2 M) \left(C_{3}-a^2 C_{5}\right)+7 \left(C_{3}-a^2 C_{5}\right)^2 \nonumber \\
&& +27 r^2 (a C_{4}+C_{2}-2 M) (-a C_{4}+C_{2}-2 M)\Bigg) \nonumber \\
&& -\frac{12}{( r^2+a^2 {\cos }^2 \theta )^3} (-a C_{4}+C_{2}-2 M) (a C_{4}+C_{2}-2 M) + \frac{k^2}{6} ~ ,\label{normRieman}
\end{eqnarray}
which might have a negative value as observed in \cite{Henry:1999rm} for Kerr-Newman metric with $k = 0$. This is so because the spacetime metric is indefinite. The norm (\ref{normRieman}) shows that the spacetime has a real ring singularity at $r=0$ and $\theta = \pi / 2$ with radius $a$.\\
Now, we would like to remark some geometrical invariants of this solution by showing the explicit expression of Komar integrals such as Komar mass and Komar angular momentum. The definition of Komar integral is given by
\begin{equation}\label{Komar}
    \mathcal{Q}=\frac{1}{N}\int_{\partial\Sigma}dS_{\mu\nu}\left(\nabla^\mu\zeta^\nu+\omega^{\mu\nu}\right),
\end{equation}
where antisymmetric tensor \(\omega^{\mu\nu}\) is the solution of
\begin{equation}\label{omega}
    \nabla_\mu\omega^{\mu\nu}=R^\nu_\rho \zeta^\rho,
\end{equation}
and \(\zeta^\nu\) is the Killing vector corresponding to the symmetry. \(N\) is any suitable normalization constant whose exact value does not have any significance. Since \eqref{Komar} is an integral on the boundary of spacelike hypersurface \(\Sigma\), we can transform the expression into a volume integral on \(\Sigma\) by using Gauss theorem. As such, we have
\begin{equation}
    \mathcal{Q}=\frac{1}{N}\int_{\partial\Sigma}dS_{\mu\nu}\nabla^\mu\zeta^\nu+\frac{1}{N}\int_{\Sigma}\sqrt{\gamma}\sigma_\mu R^\mu_\rho \zeta^\rho d^3x,
\end{equation}
with \(\sigma_\mu\) is a unit vector normal to \(\Sigma\) and \(\gamma=\text{det}[\gamma_{ij}]\) where \(\gamma_{ij}\)s are metric tensor components of hypersurface \(\Sigma\). 
Now, let us calculate the Komar mass that corresponds to the timelike Killing vector \(\zeta_{(t)}=\partial_t\). The explicit integral expression of this mass is given by
\begin{equation}
    \mathcal{M}=\frac{1}{4\pi}\int_{\partial\Sigma}dS_{01}\nabla^0\zeta^1+\frac{1}{4\pi}\int_{\Sigma}\sqrt{\gamma}\sigma_0 R^0_0 \zeta^0 d^3x.
\end{equation}
Firstly, we calculate \(\nabla^0\zeta^1\) as follows
\begin{eqnarray}
\nabla^0\zeta^1&=&g^{\mu0}\Gamma^1_{\mu\nu}\zeta^\nu\nonumber\\
&=&g^{00}\Gamma^1_{00}+g^{30}\Gamma^1_{30}\nonumber\\
&=&-\frac{2 \left(a^2+r^2\right)
   \left(\Delta_r'\left(a^2
   \cos (2 \theta )+a^2+2 r^2\right)+4
   a^2 r \Delta_\theta-4 r
   \Delta_r\right)}{\left(a^2 \cos (2
   \theta )+a^2+2 r^2\right)^3}\nonumber\\
   &=& \frac{k}{12}r-\frac{2M-C_2}{2r^2}+O(r^{-4}).
\end{eqnarray}
Since \(dS_{01}\approx -r^2\sin\theta d\theta d\varphi\) for \(r>>a\) then we have 
\begin{equation}
    \frac{1}{4\pi}\int_{\partial\Sigma}dS_{01}\nabla^0\zeta^1=M-\frac{C_2}{2}-\frac{k}{12}r^3,
\end{equation}
for the first term evaluated at large \(r\). Now, consider \(R^0_0\) that goes like
\begin{equation}
    R_0^0= \frac{1}{4}k+O(r^{-4}),
\end{equation}
at large \(r\). As such, we can expect that doing the volume integration on the second term up to an arbitrary large \(r\) gives
\begin{equation}
    \frac{1}{4\pi}\int_{\Sigma}\sqrt{\gamma}\sigma_0 R^0_0 \zeta^0 d^3x\approx \frac{k}{12} r^3.
\end{equation}
This term exactly cancels out the third term of the previous surface integral. Thus, we can conclude that Komar mass of our system is given by
\begin{equation}
    \mathcal{M}=M-\frac{C_2}{2}.
\end{equation}
As a conclusion, the constant \(C_2\) is insignificant to our solution, hence, we should take it to be zero without any loss of generality since the mass parameter, \(M\), can be redefined to absorb \(C_2\).

Now for the Komar angular momentum, the spacelike Killing vector is \(\zeta_{(\varphi)}=\partial_\varphi\). As such, the integral expression of it is given by
\begin{eqnarray}
\mathcal{J}=-\frac{1}{\pi}\int_{\partial\Sigma}dS_{01}\nabla^0\zeta^1-\frac{1}{\pi}\int_{\Sigma}\sqrt{\gamma}\sigma_0 R^0_3 \zeta^3 d^3x.
\end{eqnarray}
Again, we calculate \(\nabla^0\zeta^1\) for the new Killing vector as follows
\begin{eqnarray}
\nabla^0\zeta^1&=&g^{\mu0}\Gamma^1_{\mu\nu}\zeta^\nu\nonumber\\
&=&g^{00}\Gamma^1_{03}+g^{30}\Gamma^1_{33}\nonumber\\
&=&-\frac{a \left(a^2 (k-4
   \Lambda )+6 C_1+12
   C_4 \cos \theta -12
   C_5\right)}{12
   r}\nonumber\\&&-\frac{3 a (C_2-2
   M) \sin ^2\theta }{2 r^2}+O(r^{-3}).
\end{eqnarray}
Thus, the first term is given by
\begin{equation}
    \int_{\partial\Sigma}dS_{01}\nabla^0\zeta^1=-4 r\left[a^3\frac{k-4\Lambda}{12}+a\left(\frac{C_1}{2}-C_5\right)\right]+a\left(M-\frac{C_2}{2}\right)+O(r^{-1}).
\end{equation}
The corresponding Ricci tensor component satisfies
\begin{equation}
    R^0_3=2a\frac{C_3-a^2C_5}{r^4}\sin^2\theta+O(r^{-6}).
\end{equation}
Thus, we expect the second term goes to zero for large \(r\). As conclusion, the Komar angular momentum of this system is given by
\begin{equation}
    \mathcal{J}=a\left(M-\frac{C_2}{2}\right)-4 r\left[a^3\frac{k-4\Lambda}{12}+a\left(\frac{C_1}{2}-C_5\right)\right].
\end{equation}
We can observe that the second term in the above equation diverges as \(r\rightarrow\infty\). To avoid this problem we can choose \(C_1=2C_5\) and \(k=4\Lambda\). Thus, in order to have a physical angular momentum we need to take our solution back to the Einstein limit which gives us the following expression for Komar mass and angular momentum
\begin{equation}
    \mathcal{M}=M,~~~\mathcal{J}=aM.
\end{equation}
We can see that the parameter \(M\) and \(a\) are, indeed, similar to the mass and angular momentum parameter of classical rotating black hole. The term linear to 
\(k-4\Lambda\) and \(\frac{C_1}{2}-C_5\) that make \(\mathcal{J}\) diverges for asymptotically flat or asymptotically A-dS solution came from the statement that Ricci scalar equal to a contant, \(R=k\), which, in general, does not obey Einstein's equation.
\subsection{Regularity and Horizons}
Since an axisymmetric configurations is invariant under rotation along its azimuthal coordinate (\(SO(2)\) group on the azimuthal plane), we need to know the behaviour of the solution on the fixed points of \(SO(2)\). In four dimensional spacetime, the standard one parameter SO(2) transformation, in Cartesian coordinate bases, is given by
\begin{equation}
    O_2(\Tilde{\phi})=\begin{bmatrix}0&0&0&0\\
    0&\cos\Tilde{\phi}&\sin\Tilde{\phi}&0\\
    0&-\sin\Tilde{\phi}&\cos\Tilde{\phi}&0\\0&0&0&0
    \end{bmatrix},
\end{equation}
with \(\Tilde{\phi}\) is the parameter of transformation. Fixed points, \(X\), for this symmetry group satisfies \(X=O_2X\). The solutions for such equation is \begin{equation}
    X=\begin{bmatrix}t\\0\\0\\z\end{bmatrix},
\end{equation}
with \(t\) and \(z\) are arbitrary. Transforming these points into spherical coordinate system gives us arbitrary \(t\), \(r\), and \(\phi\) with polar coordinate must be fixed at \(\theta=0\) or \(\theta=\pi\). This set of fixed points is known as the axes of rotation.

Evaluating \eqref{axisymmetricsol} at \(\theta=\pi\) or \(\theta=0\) gives
\begin{equation}
ds^2= -e^{2\nu } dt^2 + e^{2\psi }{\left(d\varphi -\omega dt\right)}^2 + e^{2{\mu }_2} dr^2 , 
\label{axes} 
\end{equation} 
with
\begin{eqnarray}
e^{2{\mu }_2}&=&\frac{(r^2+a^2)}{\Delta_r}\nonumber\\
e^{2(\psi+\nu)} &=& {\Delta }_r {\Delta }_{\theta }   ~ , \nonumber\\
e^{2\psi } &=&   \left(r^2+a^2\right) \Delta_{\theta} ~ , \\
e^{2\nu } &=&  \frac{  \Delta_r }{  \left(r^2+a^2\right) } ~,  \nonumber\\
\omega &=& \frac{ a} { \left(r^2+a^2\right) }  ~ , \nonumber
\end{eqnarray}
 ${\Delta }_{\theta } $ is now given by
\begin{eqnarray}
\Delta_{\theta} &=& \frac{\Lambda }{3} a^2 - \frac{k}{12} a^2-\frac{1}{2}C_1\pm C_4+C_5   ~ . \nonumber  
\end{eqnarray}
Transforming \(r\) to the Cartesian coordinate  gives the following metric
\begin{equation}
    ds^2=-\left(\frac{\Delta_r-a^2\Delta_\theta}{z^2+a^2}\right)dt^2+\left(\frac{z^2+a^2}{\Delta_r}\right)dz^2.
\end{equation}
The resulting two dimensional manifold on the axis of rotation have similar coordinate singularity (Back hole horizon) with the full rotating solution but the true singularity is only one, in contrast with the full solution that possess ring singularity as well. We can observe that taking \(a=0\), implying zero angular momentum, gives us a Riessner-Nordstrom-de Sitter solution with extra term containing \(C_1\) and \(C_2\). 

There is a special region where the metric changes its signature, that is when \(\Delta_r<a^2\Delta_\theta\), Thus for every \(z\) that solves \(\mathcal{G}(z)>0\), where 
\begin{eqnarray}
    \mathcal{G}(z)&\equiv& a^2\Delta_\theta-\Delta_r\nonumber\\
    &=&\frac{\Lambda }{3} a^2\left(a^2+z^2\right) - \frac{k}{12} \left(a^4-z^4\right)-\left(z^2+a^2\right)+2Mz\nonumber\\&&-\frac{1}{2}C_1\left(a^2+z^2\right)-C_2 z - C_3 \pm C_4a^2+C_5a^2 \nonumber\\
\end{eqnarray}
the spacetime is locally Euclidean. This region is located near black hole horizon where \(\Delta_r\) is close to zero. The existence of these locally Euclidean is a problem because the boundary between locally Minkowskian and locally Euclidean region is irregular. We should again argue that this irregularity comes from the fact that our formulation is more general than Einstein's equation (which is always assumed to be locally Minkowskian everywhere) and such extension allows irregularities to show up in our solution. There is actually some ways to proof that this irregularity is not pathological, for example, by showing that these locally Euclidean regions are hidden behind blackhole horizons, which will be discussed below.

\indent Consider the mean extrinsic curvature of a two-dimensional spatial hypersurface constructed from orthonormal timelike vector that is orthogonal to the three dimensional hypersurface and an outward (radial) pointing vector for this spacetime, given by
\begin{equation}\label{MeanCurv}
    \mathcal{K}=\frac{\Delta_r \left(4 r \left(a^2+r^2\right) \Delta_\theta-a^2 \sin ^4\theta  \Delta_r'\right)}{2 \sqrt{2} \left(a^2 \cos ^2\theta +r^2\right) \sqrt{\frac{\Delta_r}{a^2 \cos (2 \theta
   )+a^2+2 r^2}} \left(\left(a^2+r^2\right)^2 \Delta_\theta -a^2 \sin ^4\theta  \Delta_r\right)}.
\end{equation}
The apparent horizons are solutions of \(\mathcal{K}=0\) and we can directly see that the event horizons, where \(\Delta_r=0\), coincides with some of the apparent horizons. 

Firstly, let us consider the locations of event horizons by solving \(\Delta_r=0\). The problem of finding event horizons in this spacetime can be reduced to a problem of solving quartic equation given by
\begin{equation}
    - {\mathcal A} r^4+ {\mathcal B} r^2 + {\mathcal C} r + {\mathcal D} = 0 ~ , \label{thePolynom}
\end{equation}
where we have defined
\begin{eqnarray}
{\mathcal A} &\equiv& \frac{k}{12} ~ , \nonumber \\
{\mathcal B} &\equiv& 1+\frac{C_{1}}{2}-\frac{a^2 \Lambda }{3} ~ , \nonumber \\
{\mathcal C} &\equiv& -2 M ~ , \\
{\mathcal D} &\equiv& a^2+C_{3} ~. \nonumber
\end{eqnarray}
The quartic equation \eqref{thePolynom} has four possible solutions with one solution is a definite negative, hence, we are left with only three possible solutions for the horizons.

The two smaller solutions should be considered as the blackhole horizons that is given by
\begin{eqnarray}
    r_+=\frac{1}{2}\sqrt{\phi_1}-\frac{1}{2}\sqrt{\phi_2+\frac{2\mathcal{C}}{\mathcal{A}\sqrt{\phi_3}}}\\
    r_-=-\frac{1}{2}\sqrt{\phi_1}+\frac{1}{2}\sqrt{\phi_2-\frac{2\mathcal{C}}{\mathcal{A}\sqrt{\phi_3}}}
\end{eqnarray}
such that \(r_+\geq r_-\) and we have defined 
\begin{eqnarray}
    \phi_1&\equiv& \frac{2 \mathcal{B}}{3 \mathcal{A}}-\frac{\sqrt[3]{\sqrt{\left(72 \mathcal{A} \mathcal{B} \mathcal{D}-27 \mathcal{A} \mathcal{C}^2+2 \mathcal{B}^3\right)^2-4 \left(\mathcal{B}^2-12 \mathcal{A} \mathcal{D}\right)^3}+72 \mathcal{A} \mathcal{B} \mathcal{D}-27 \mathcal{A} \mathcal{C}^2+2 \mathcal{B}^3}}{3
   \sqrt[3]{2} \mathcal{A}}\nonumber\\&&-\frac{\sqrt[3]{2} \left(\mathcal{B}^2-12 \mathcal{A} \mathcal{D}\right)}{3 \mathcal{A} \sqrt[3]{\sqrt{\left(72 \mathcal{A} \mathcal{B} \mathcal{D}-27 \mathcal{A} \mathcal{C}^2+2 \mathcal{B}^3\right)^2-4
   \left(\mathcal{B}^2-12 \mathcal{A} \mathcal{D}\right)^3}+72 \mathcal{A} \mathcal{B} \mathcal{D}-27 \mathcal{A} \mathcal{C}^2+2 \mathcal{B}^3}}~,\nonumber\\
   \\ \phi_2&\equiv&\frac{4 B}{3 A}+\frac{\sqrt[3]{\sqrt{\left(72 \mathcal{A} \mathcal{B} \mathcal{D}-27 \mathcal{A} \mathcal{C}^2+2 \mathcal{B}^3\right)^2-4 \left(\mathcal{B}^2-12 \mathcal{A} \mathcal{D}\right)^3}+72 \mathcal{A} \mathcal{B} \mathcal{D}-27 \mathcal{A} \mathcal{C}^2+2 \mathcal{B}^3}}{3
   \sqrt[3]{2} \mathcal{A}}\nonumber\\&&
   +\frac{\sqrt[3]{2} \left(\mathcal{B}^2-12 \mathcal{A} \mathcal{D}\right)}{3 \mathcal{A} \sqrt[3]{\sqrt{\left(72 \mathcal{A} \mathcal{B} \mathcal{D}-27 \mathcal{A} \mathcal{C}^2+2 \mathcal{B}^3\right)^2-4
   \left(\mathcal{B}^2-12 \mathcal{A} \mathcal{D}\right)^3}+72 \mathcal{A} \mathcal{B} \mathcal{D}-27 \mathcal{A} \mathcal{C}^2+2 \mathcal{B}^3}} ~,\nonumber\\ \\
   \phi_3&\equiv& \frac{2 \mathcal{B}}{3 \mathcal{A}}-\frac{3
   \sqrt[3]{2} \mathcal{A}}{\sqrt[3]{\sqrt{\left(72 \mathcal{A} \mathcal{B} \mathcal{D}-27 \mathcal{A} \mathcal{C}^2+2 \mathcal{B}^3\right)^2-4 \left(\mathcal{B}^2-12 \mathcal{A} \mathcal{D}\right)^3}+72 \mathcal{A} \mathcal{B} \mathcal{D}-27 \mathcal{A} \mathcal{C}^2+2 \mathcal{B}^3}}\nonumber\\&&-\frac{3 \mathcal{A} \sqrt[3]{\sqrt{\left(72 \mathcal{A} \mathcal{B} \mathcal{D}-27 \mathcal{A} \mathcal{C}^2+2 \mathcal{B}^3\right)^2-4
   \left(\mathcal{B}^2-12 \mathcal{A} \mathcal{D}\right)^3}+72 \mathcal{A} \mathcal{B} \mathcal{D}-27 \mathcal{A} \mathcal{C}^2+2 \mathcal{B}^3}}{\sqrt[3]{2} \left(\mathcal{B}^2-12 \mathcal{A} \mathcal{D}\right)}~.\nonumber\\
\end{eqnarray}
The biggest solution of \eqref{thePolynom} is the cosmological horizon that is given by
\begin{eqnarray}
    r_c=\frac{1}{2}\sqrt{\phi_1}+\frac{1}{2}\sqrt{\phi_2+\frac{2\mathcal{C}}{\mathcal{A}\sqrt{\phi_3}}}.
\end{eqnarray}
From here, we can see that the physical region lies in \(r_+<r<r_c\) where the metric is timelike. The region \(r_-\leq r\leq r_+\) and \(r\geq r_c\) have spacelike metric, thus, the qualitative features of this spacetime related to the location of timelike region is similar to the one we found in Einstein limit, except the fact that this solution possess some locally Euclidean regions. Since these locally Euclidean regions are located behind the horizons, then this region is not pathological and do not have any physical significance.

The number of apparent horizons of this spacetime is actually more than the number of event horizons since taking \eqref{MeanCurv} equal to zero can also be done by taking
\begin{equation}
    4 r \left(a^2+r^2\right) \Delta_\theta-a^2 \sin ^4\theta  \Delta_r'=0,
\end{equation}
that possess, at most, three different solutions for \(r\). This proves that there exist trapped regions that does not coincide with event horizons at their boundaries. 
Since we have two black hole horizons, it is interesting to consider a 'critical' case where the 'discriminant' of the quartic polynomial (\ref{thePolynom}) vanishes, namely
\begin{equation}\label{Discrim}
-256 {\mathcal A}^3 {\mathcal D}^3-128 {\mathcal A}^2 {\mathcal B}^2 {\mathcal D}^2+144 {\mathcal A}^2 B {\mathcal C}^2 {\mathcal D} - 27 {\mathcal A}^2 {\mathcal C}^4-16 {\mathcal A} {\mathcal B}^4 {\mathcal D} + 4 {\mathcal A} {\mathcal B}^3 {\mathcal C}^2 = 0 ~ ,
\end{equation}
with $k \ne 0$. The roots of  (\ref{Discrim}) have the form
\begin{equation}
 2 M =  \sqrt{\frac{2}{27}} \sqrt{\frac{{\mathcal B}^3}{{\mathcal A}}+36 {\mathcal B} {\mathcal D} \pm \frac{\sqrt{\left({\mathcal B}^2-12 {\mathcal A} {\mathcal D}\right)^3}}{{\mathcal A}}} ~ ,
\end{equation}
which gives a  'critical' mass of  a black hole describing a situation where some roots of (\ref{thePolynom}) coincide for $k > 0$. In addition, the term inside the square root must be positive in order to have a physical solution.\\
\indent Generally, the metric described in Theorem 1  may not be related to  Einstein general relativity since our method described above does not use the notion of energy-momentum tensor. To make a contact with general relativity, we could simply set some constants, for example,  namely 
\begin{equation}  \label{KerNewEin}
C_1 =  C_2 = C_4 = C_5 = 0 ~ , ~ C_3 = q^2 + g^2 ~ , 
\end{equation}
with $k = 4\Lambda$ where $q$ and $g$ are electric and magnetic charges, respectively. This setup gives the Kerr-Newman-Einstein metric  describing a dyonic rotating black hole with non-zero cosmological constant \cite{Setare:2003jc}. For $\Lambda > 0$, the  'critical' mass of  a black hole in this case is simpy given by
\begin{eqnarray}\label{special}
M &=& \sqrt{\frac{1}{54}} \left[ \frac{3}{\Lambda } \left(1-\frac{a^2 \Lambda }{3}\right)^3+36 \left(1-\frac{a^2 \Lambda }{3}\right) \left(a^2 + q^2 + g^2 \right) \right. \nonumber \\
&& \left. - \frac{3}{\Lambda } \left(\left(1-\frac{a^2 \Lambda }{3}\right)^2-4 \Lambda \left(a^2 + q^2 + g^2 \right)\right)^{3/2} \right]^{1/2}
\end{eqnarray}
 where the inner horizon and the event horizon coincide. In this critical case, the inner time-like region \(0\leq r \leq R_-\) and the physical region \(r_+\leq r \leq r_c\) are connected, which leads to a naked singularity at the origin.

\subsection{Numerical Results}\label{Num}
\begin{figure}
\centering
\includegraphics[width=0.8\textwidth]{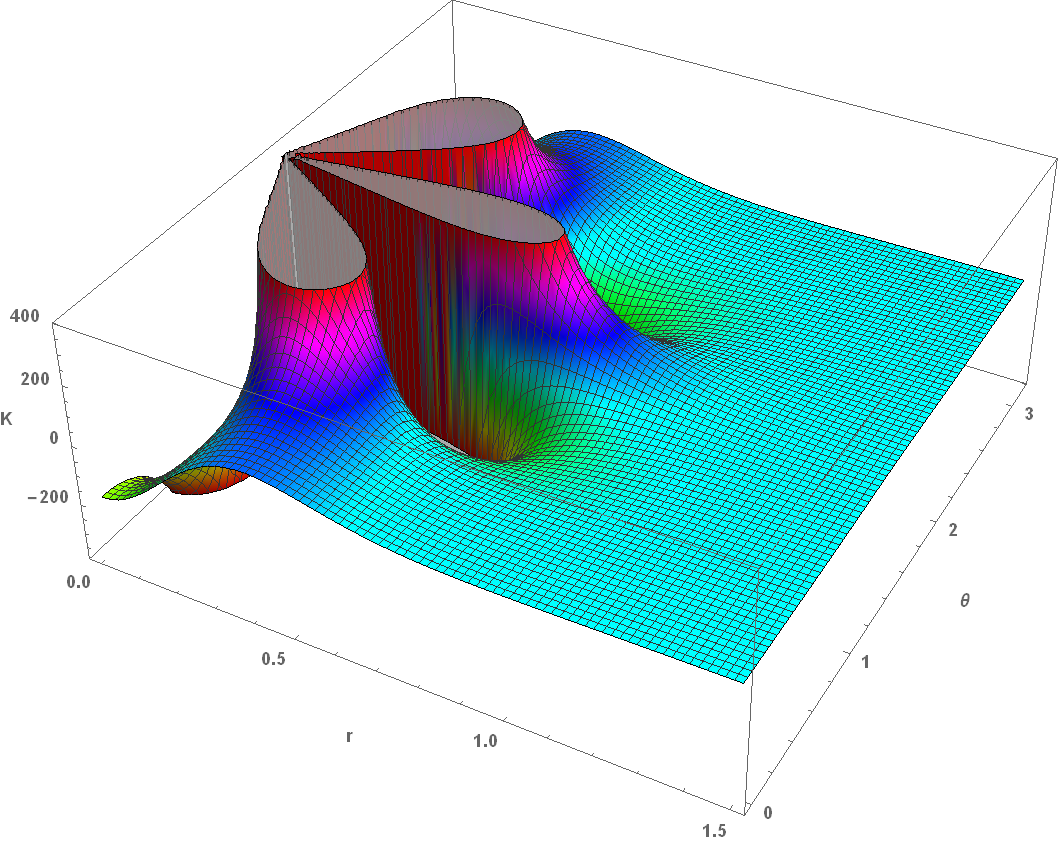}
\caption{Plot of Kretschmann scalar as function of coordinates for Kerr solution with $a=0.8$.}\label{kretschmannkerrplot}
\end{figure}

\begin{figure}
\centering
\includegraphics[width=0.8\textwidth]{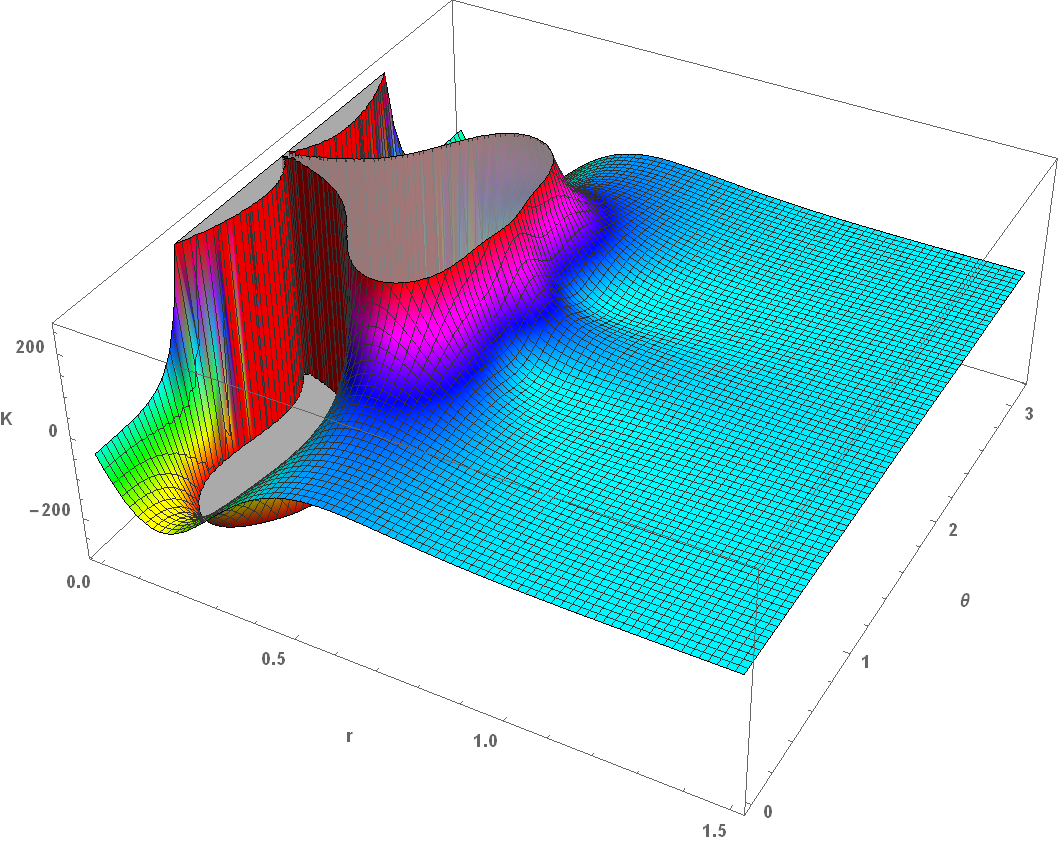}
\caption{Plot of Kretschmann scalar as function of coordinates for Kerr-Newmann solution with $a=0.8$ and $C_3 = q^2+g^2 = 0.64$.}\label{kretschmannkerrnewmanplot}
\end{figure}

With the Kretschmann \eqref{Kretschmann} at hand, we can identify the true singular points within our metric solutions because Kretschmann scalar become singular at singular point. Some black hole classes might have different true singular point which depends on their parameters. We conclude the results  as follows.
\begin{enumerate}
\item For every static blackhole solutions, $r=0$ is the true singularity
\item For every stationary axisymmetric blackhole solutions, $r^2+a^2\cos^2\theta=0$ is the true singularity. 
\end{enumerate}
To see this clearer, we plot some profile of the Kretschmann scalar for Kerr and Kerr-Newman blackholes by tweaking our parameters to reproduce those three blackhole solutions. These plots are given in Figure (\ref{kretschmannkerrplot}-\ref{kretschmannkerrnewmanplot}) with \(M=1\) for Schwarzshild solution, \(a=0.8\) for Kerr solution, and $a=0.8$ and $C_3 = q^2+g^2 = 0.64$ for Kerr-Newmann solution.

\begin{figure}
\centering
\includegraphics[width=0.4\textwidth]{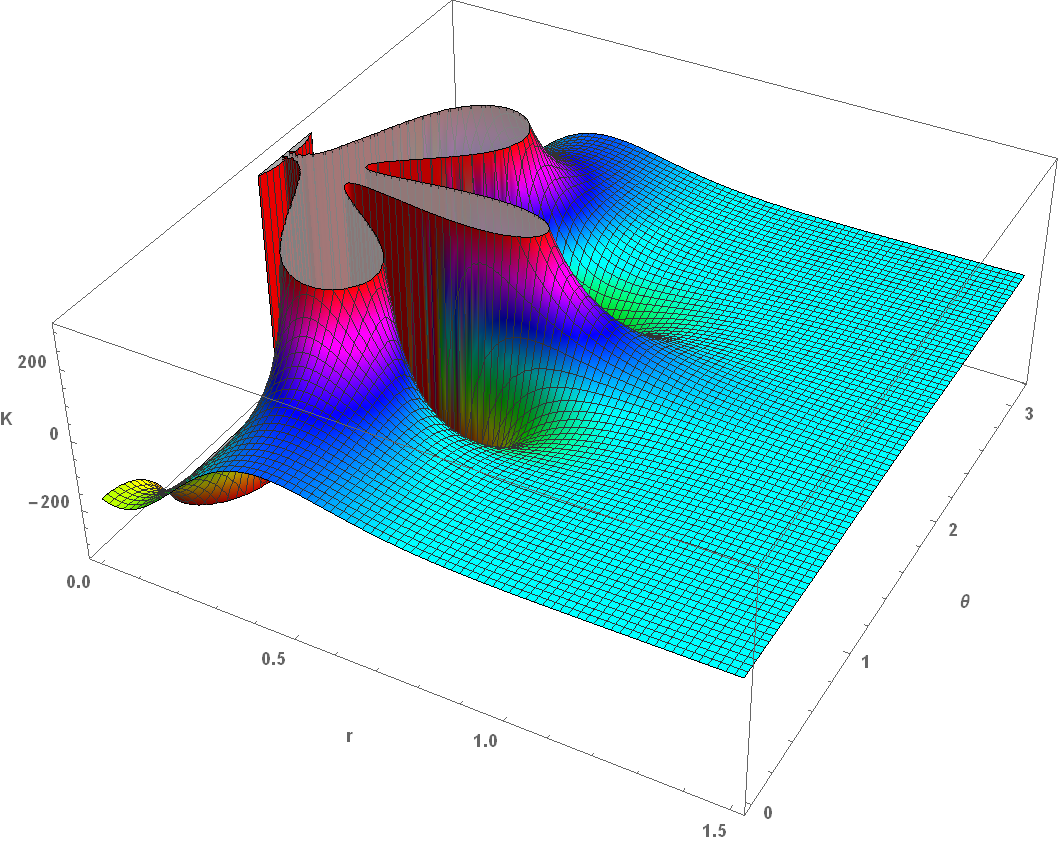}
\includegraphics[width=0.4\textwidth]{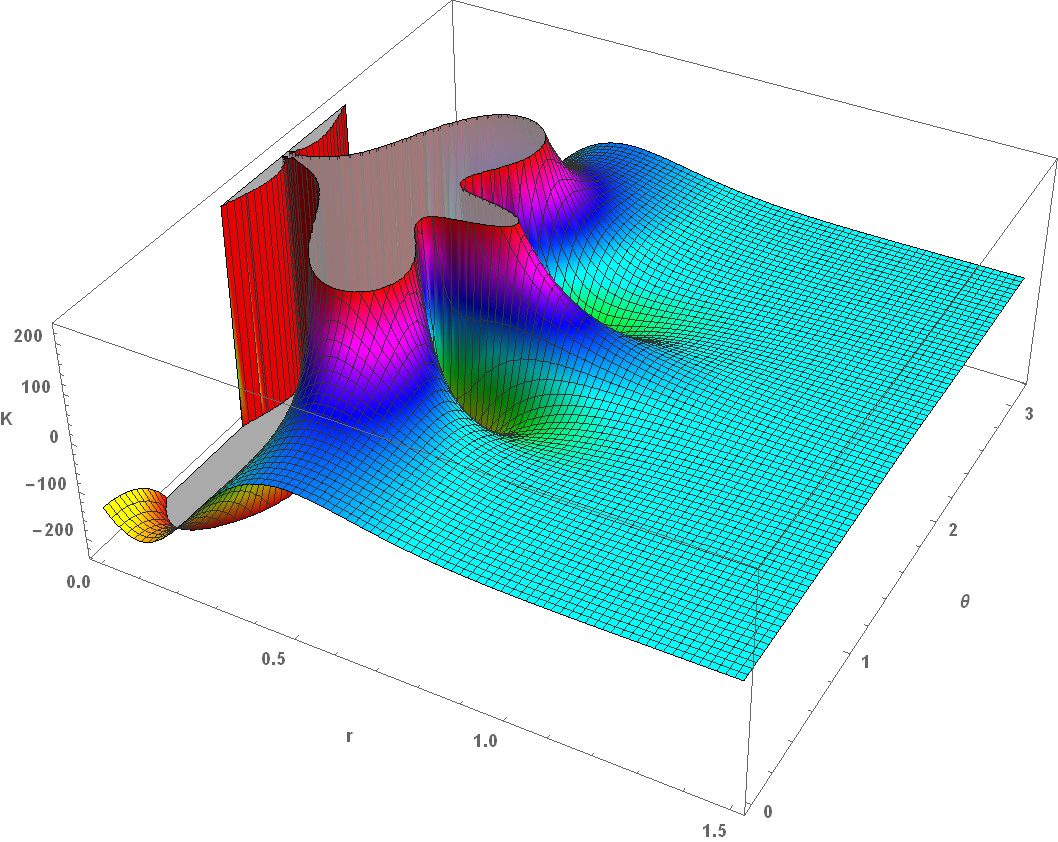}
\includegraphics[width=0.4\textwidth]{kretschmann-kerr-newmann_a=0_8-C3=0_64.png}
\includegraphics[width=0.4\textwidth]{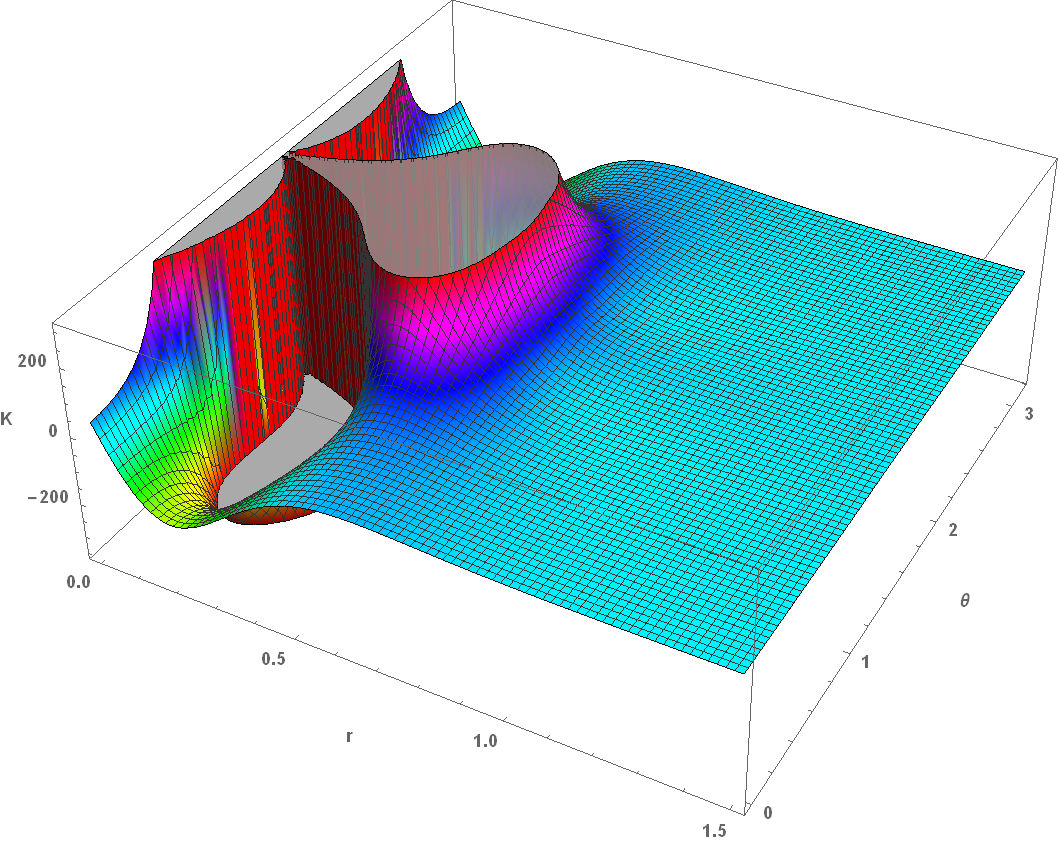}
\caption{Demonstration on connection between Kerr and Kerr-Newman solution by increasing blackhole charge with $a=0.8$ for $C_3 = q^2+g^2 = \{0.16,\ 0.36,\ 0.64,\ 0.81\}$.}\label{kretschmannkerrnewmanchargecomparison}
\end{figure}

 From figure \ref{kretschmannkerrplot} and figure \ref{kretschmannkerrnewmanplot} we observe  the real ring singularity at $r=0$ and $\theta = \pi / 2$ with radius $a$ which are mentioned in previous section. Locations of singular points does not affected by asymptotic structure of spacetime, hence the value of constant Ricci scalar does not alter the singularity. We can see that for stationary cases, there exist a region where Kretschmann scalar takes negative value near singularity. One interesting fact is the singularity can be avoided for inward radial motion through the blackhole north and south poles. This indicate a fundamentally different structure between those spacetime manifolds. The Kerr solution and Kerr-Newman solution is connected i.e. taking chargeless limit of Kerr-Newman solution reproduces Kerr solution, which is demonstrated in figure \ref{kretschmannkerrnewmanchargecomparison}. 

\begin{figure}\label{regions}
\centering
\includegraphics[scale=0.15]{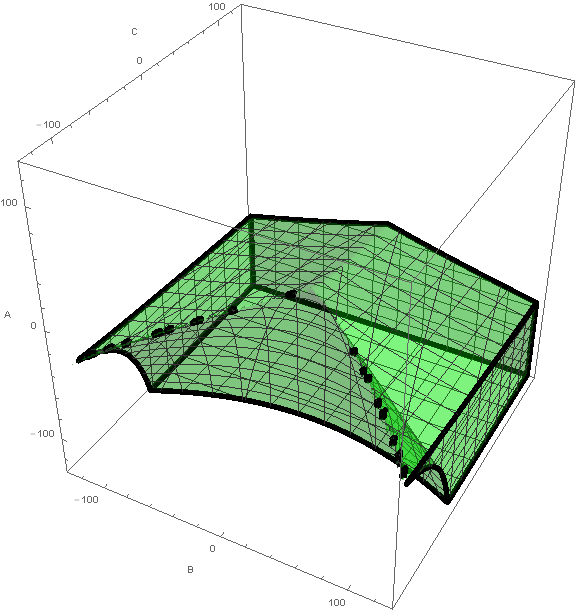}
\includegraphics[scale=0.15]{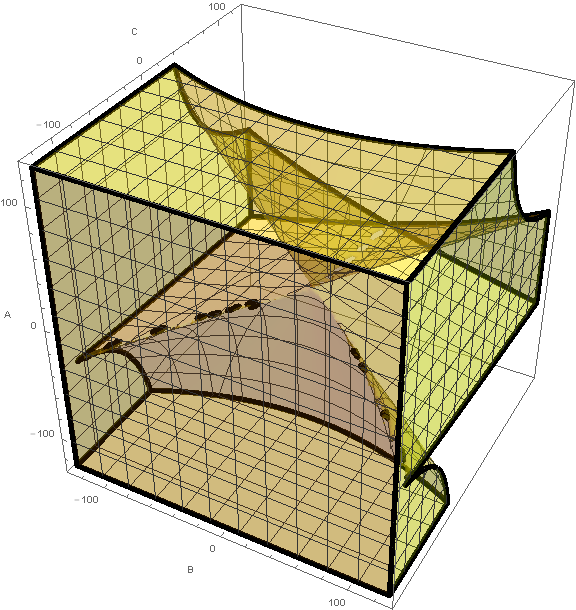}
\includegraphics[scale=0.15]{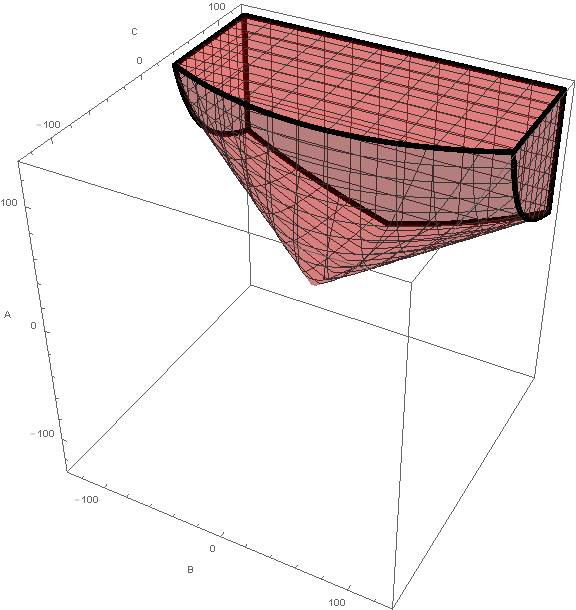}
\caption{three parameter space spanned by \(A,B,C\) with Region I (left) has real 4 solutions, region II (middle) has 2 real solutions, and region III (right) has no real solution.}
\end{figure}

\begin{figure}
\centering
\includegraphics[scale=0.2]{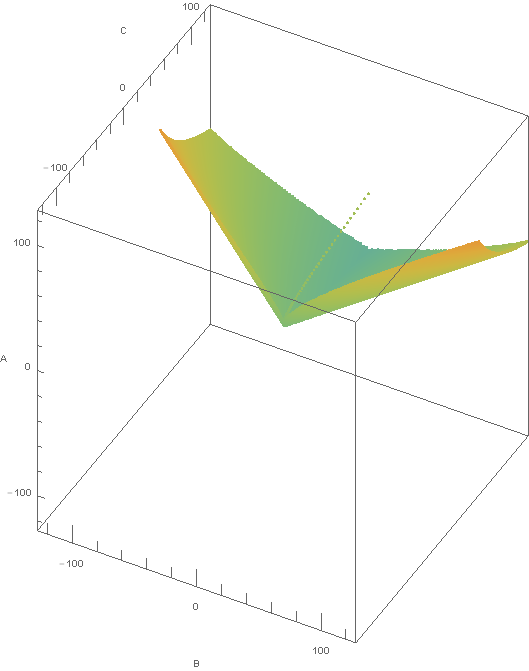}
\includegraphics[scale=0.2]{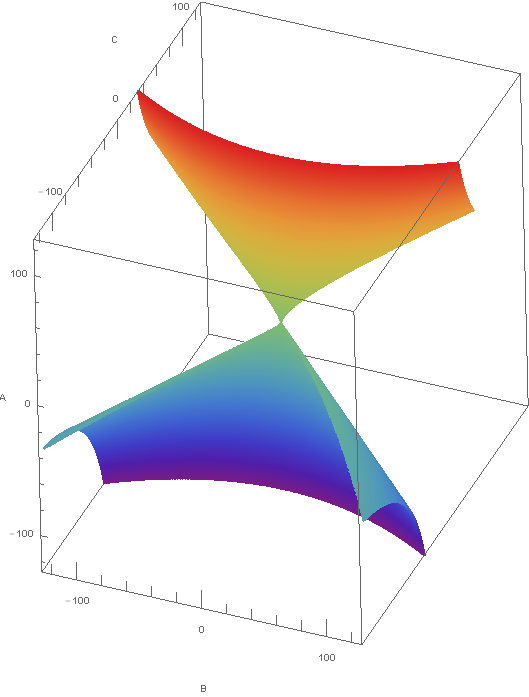}
\includegraphics[scale=0.2]{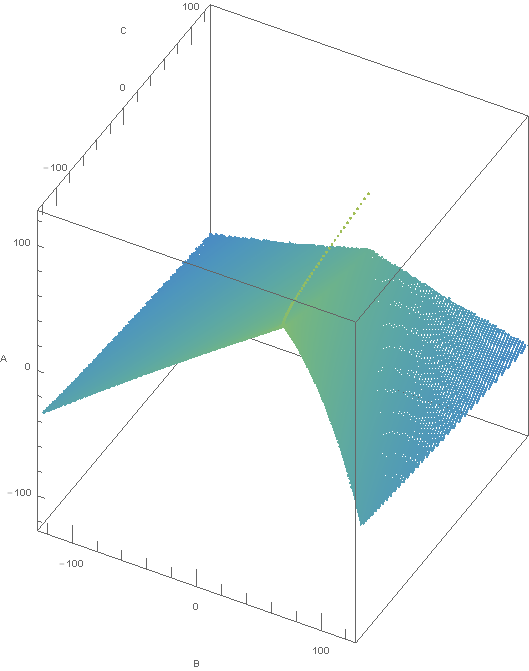}
\caption{The three surfaces of $D = 0$. Each surface represents $C_{\text{I}}$ (left), $C_{\text{II}}$ (middle), and $C_{\text{III}}$ (right).}
\label{discriminant-c1}
\end{figure}

The special results in \eqref{special} for critical blackhole in which all horizons coincide can be generalized, where parameters other than \(M\) \(\Lambda\) and \(a\) are taken into account. Firstly, we introduce another reduced parameter \(A, B, C\) for equation \eqref{thePolynom} which satisfy \(A=\frac{\mathcal{B}}{\mathcal{A}}\), \(B=\frac{\mathcal{C}}{\mathcal{A}}\), and \(C=\frac{\mathcal{D}}{\mathcal{A}}\). The corresponding discriminant equation, \(D=0\), becomes
\begin{equation}\label{disc}
    16 A^4 C-4 A^3 B^2-128 A^2 C^2+144 A B^2 C-27 B^4+256 C^3=0
\end{equation}
Equation \eqref{disc} represents surfaces in three dimensional space spanned by \(A,B,C\). These surfaces divides the space into three regions, namely: Region I which has 4 real solutions, region II which has 2 real solutions, and region III which does not have real solution (see figure \ref{regions}). 
The surfaces that emerge from discriminant equation can be divided into three independent surfaces. We can identify the three surfaces by solving equation \eqref{disc} as cubic equation for \(C\) which gives three solutions, namely
\begin{eqnarray}
C_{\text{I}} &=& \frac{A^2}{6}+\frac{1}{24} \sqrt[3]{-8 A^6-540 A^3 B^2+3 \sqrt{3} \sqrt{B^2 \left(8 A^3+27 B^2\right)^3}+729 B^4} \nonumber \\
&& +\frac{A^4-27 A B^2}{6 \sqrt[3]{-8 A^6-540 A^3 B^2+3 \sqrt{3} \sqrt{B^2 \left(8 A^3+27 B^2\right)^3}+729 B^4}} \\
C_{\text{II}} &=& \frac{A^2}{6}+\frac{1}{48} i \left(\sqrt{3}+i\right) \sqrt[3]{-8 A^6-540 A^3 B^2+3 \sqrt{3} \sqrt{B^2 \left(8 A^3+27 B^2\right)^3}+729 B^4} \nonumber \\
&& +\frac{\left(i \sqrt{3}+1\right) \left(27 A B^2-A^4\right)}{12 \sqrt[3]{-8 A^6-540 A^3 B^2+3 \sqrt{3} \sqrt{B^2 \left(8 A^3+27 B^2\right)^3}+729 B^4}}
\end{eqnarray}
\begin{eqnarray}
C_{\text{III}} &=& \frac{A^2}{6}-\frac{1}{48} i \left(\sqrt{3}-i\right) \sqrt[3]{-8 A^6-540 A^3 B^2+3 \sqrt{3} \sqrt{B^2 \left(8 A^3+27 B^2\right)^3}+729 B^4} \nonumber \\
&& -\frac{\left(i \sqrt{3}-1\right) \left(27 A B^2-A^4\right)}{12 \sqrt[3]{-8 A^6-540 A^3 B^2+3 \sqrt{3} \sqrt{B^2 \left(8 A^3+27 B^2\right)^3}+729 B^4}}
\end{eqnarray}
Each solutions represents each independent surface which are given in figure \ref{discriminant-c1}. 
By identifying all the curves which arises from intersections between the surfaces, we conclude that there are five different characteristics of critical point

\begin{tabular}{ llll } 
1. & quadruple point & : & $A = B = C = 0$\\
2. & triple point & : & $A = 2 i \sqrt{3} \sqrt{C}, \qquad B = \pm \frac{8 \sqrt[4]{-1} C^{3/4}}{3^{3/4}}, \qquad C < 0$ \\
3. & 2 double points & : & $A = -2 \sqrt{C}, \qquad B = 0, \qquad C>0$ \\
4. & double points and 2 real & : & $A<0$, $B$ indetermined, and $C = C_{\text{II}}$ or $C = C_{\text{III}}$ \\
5. & double points and 2 complex & : & $A$ and $B$ indetermined, and $C = C_{\text{I}}$ or $C = \text{upper~} C_{\text{II}}$ \\
\end{tabular}

Because we know that parameters \(A,B\) and \(C\) are functions of blackhole parameters \(M, k, \Lambda, a,\) and \(C_i\)'s, then the five characteristics mentioned above lead to five sets of relations which represents different classes of critical blackhole solutions.

\section{Conclusion}\label{conclusions}
We have shown in Theorem \ref{theor} that there exist a family of constant curvature axisymmetric stationary spacetimes characterized by the metric solution \eqref{sol}. The newfound solution generalized well-known solutions such as Kerr and Kerr-Newman solution by introducing four new parameters (see Table \ref{bhcomparison}).
\begin{table}[h]
\caption{ Comparison of some well-known blackhole solutions. ( *Produced From Einstein limit which also has constant scalar curvature, but are constrained by Einstein equation) }
\label{bhcomparison}
\renewcommand{\arraystretch}{1.6}
\begin{tabular}{ c c c c c c c c c c}
 ~& $M$ & $a$ & $\Lambda$ & $k$ & $C_1$ &  $C_3$ & $C_4$ & $C_5$ \\
 \hline
 Schwarzschild & \checkmark & $0$ & $0$ &  $0$ & $0$ & $0$ & $0$ & $0$ \\ \hline
 Kerr & \checkmark & \checkmark & $0$ & $0$ & $0$ & $0$ & $0$ & $0$ \\ \hline
 Reissner-Nordstr{\"o}m & \checkmark & $0$ & $0$ & $0$ & $0$ &  $q^2+g^2$ & $0$ & $0$ \\ \hline
 Kerr-Newman & \checkmark & \checkmark & $0$ & $0$ & $0$ & $q^2+g^2$ & $0$ & $0$ \\ \hline
 Schwarzschild-(A)dS & \checkmark & $0$ & \checkmark & $4 \Lambda$ & $0$ & $0$ &  $0$ & $0$ \\ \hline
 Kerr-(A)dS & \checkmark & \checkmark & \checkmark & $4 \Lambda$ & $0$ & $0$ & $0$ & $0$ \\ \hline
 Reissner-Nordstr{\"o}m-(A)dS & \checkmark & $0$ & \checkmark & $4 \Lambda$ & $0$ &  $q^2+g^2$ & $0$ & $0$ \\ \hline
 Kerr-Newman-(A)dS & \checkmark & \checkmark & \checkmark & $4 \Lambda$ & $0$ & $q^2+g^2$ & $0$ & $0$ \\ \hline
 Kerr-Newman-(A)dS* & \checkmark & \checkmark & \checkmark & $4 \Lambda$ & \checkmark & $a^2 C_5$ & \checkmark & \checkmark \\ \hline
 This paper & \checkmark & \checkmark & \checkmark & \checkmark & \checkmark & \checkmark & \checkmark & \checkmark \\ \hline
\end{tabular}
\renewcommand{\arraystretch}{1}
\end{table}
The true singularities are identified by utilizing Kretschmann scalar as given in Figure \ref{kretschmannkerrplot}-\ref{kretschmannkerrnewmanplot} and it is found that static and stationary solutions have different Kretschmann scalar  profile characteristic. We found that  four coordinate singularities are generally present and there are five different classes of critical conditions of blackhole parameters where the horizons coincide. 

\appendix

\section{SPACETIME CONVENTION}
In this section we collect some spacetime quantities which are useful for the analysis in the paper. \\

Christoffel symbol: 
\begin{equation} 
{{\Gamma }^{\lambda }}_{\mu \nu }=\frac{1}{2} g^{\rho \sigma }\left( \partial_{\mu}g_{\rho \nu }  +  \partial_{\nu}g_{\rho \mu } - \partial_{\rho} g_{\mu \nu } \right)  ~ . 
\end{equation} 
\\
\indent Riemann curvature tensor:
\begin{equation} 
- R^{\rho }_{\mu \nu \sigma }={\partial }_{\sigma }{{\Gamma }^{\rho }}_{\mu \nu }-{\partial }_{\nu }{{\Gamma }^{\rho }}_{\mu \sigma }+{{\Gamma }^{\lambda }}_{\mu \nu }{{\Gamma }^{\rho }}_{\lambda \sigma }-{{\Gamma }^{\lambda }}_{\mu \sigma }{{\Gamma }^{\rho }}_{\lambda \nu } ~ . 
\end{equation} 
\\
\indent Ricci tensor:
\begin{equation} 
R_{\mu \nu }=R^{\rho }_{\mu \rho v}={\partial }_{\rho }{{\Gamma }^{\rho }}_{\mu \nu }-{\partial }_{\nu }{{\Gamma }^{\rho }}_{\mu \rho }+{{\Gamma }^{\lambda }}_{\mu \nu }{{\Gamma }^{\rho }}_{\lambda \rho }-{{\Gamma }^{\lambda }}_{\mu \rho }{{\Gamma }^{\rho }}_{\lambda \nu } ~ .
\end{equation} 
\\
\indent Ricci scalar:
\begin{equation}  
R=g^{\mu \nu }R_{\mu \nu }  ~ .
\end{equation} 

\section{Event Horizon Kerr-deSitter and Kerr-Newman-deSitter Solutions}
\begin{figure}
\centering
\includegraphics[width=0.6\textwidth]{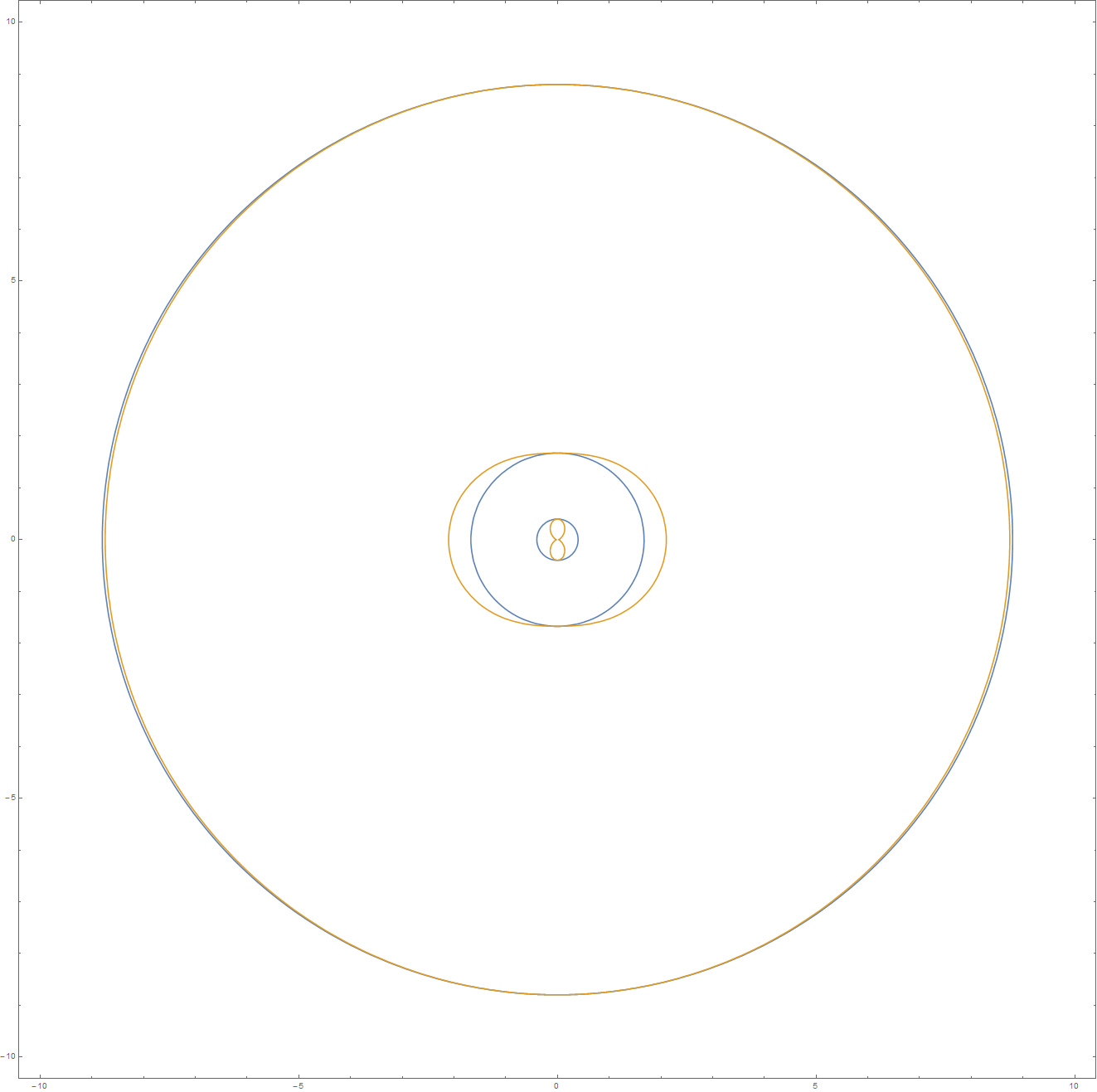}
\caption{ Event horizon (blue) and stationary limit (orange) of Kerr-de Sitter blackhole with $M=1$, $a=0.8$, and $\Lambda = 0.03$.}
\label{kdshorizonfig}
\end{figure}

\begin{figure}
\centering
\includegraphics[width=0.6\textwidth]{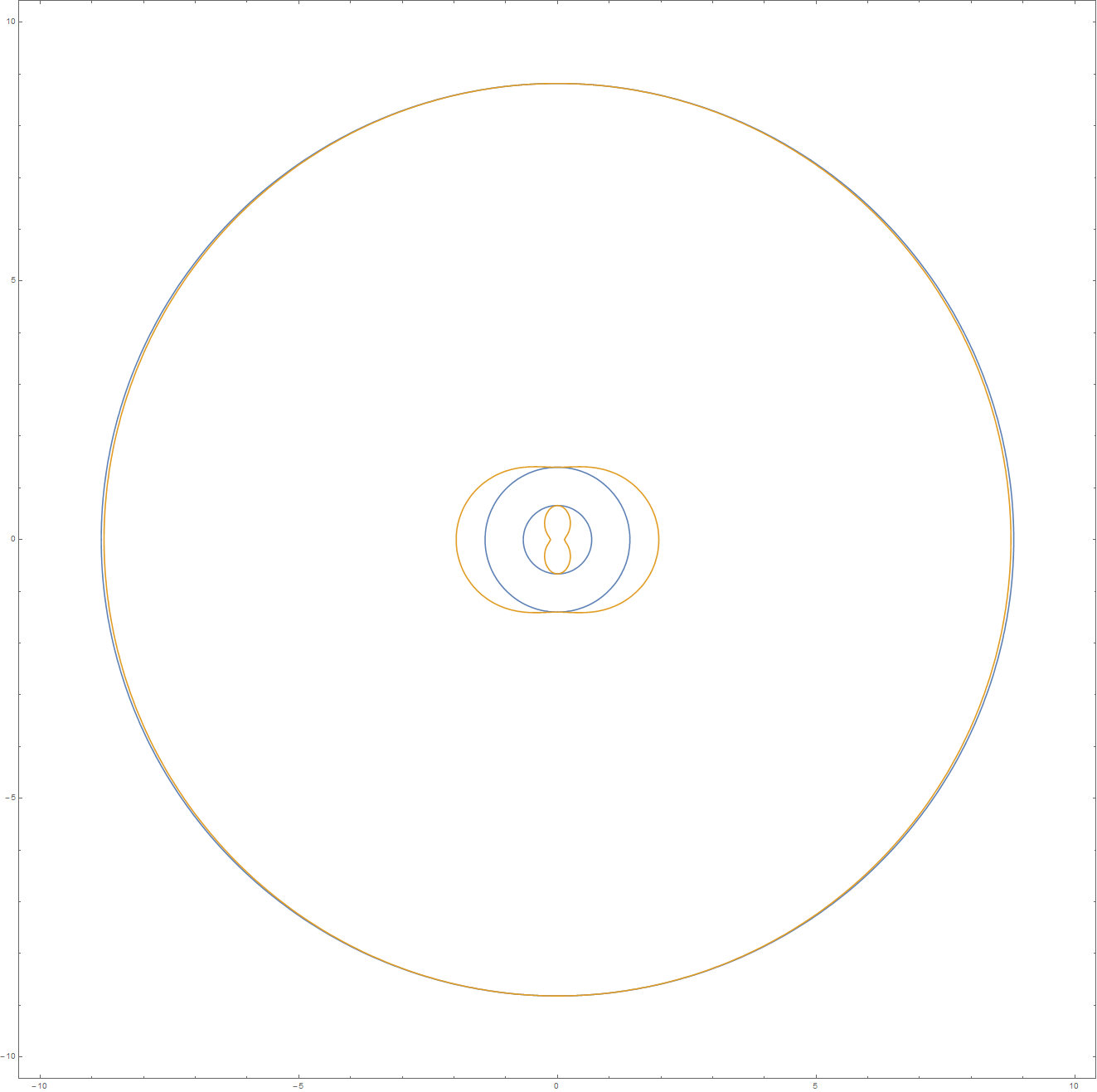}
\caption{ Event horizon (blue) and stationary limit (orange) of Kerr-Newman-de Sitter blackhole with $M=1$, $a=0.8$, $C_3 = q^2+g^2 = 0.25$ and $\Lambda = 0.03$.}
\label{kndshorizonfig}
\end{figure}

In order to visualize clearer the shape of the horizon that occurs, we give here two examples of event horizons for Kerr-de Sitter and Kerr-Newman-de Sitter black holes, shown in Figure \ref{kdshorizonfig} and Figure \ref{kndshorizonfig}. The value of $\Lambda$ is chosen to be large enough such that cosmological horizon can be observed.

\newpage

\section*{Acknowledgments}

The work in this paper is supported by Riset KK ITB, Riset ITB, and PDUPT Kemendikbudristekdikti-ITB.

\end{document}